\documentclass[prb,
twocolumn,
superscriptaddress,showpacs,amsmath,amssymb]{revtex4}

\begin{document}

\author{G.E.~Volovik}
\affiliation{Low Temperature Laboratory, Aalto University,  P.O. Box 15100, FI-00076 Aalto, Finland}
\affiliation{Landau Institute for Theoretical Physics, acad. Semyonov av., 1a, 142432,
Chernogolovka, Russia}

\title{Painlev\'e-Gullstrand coordinates for Schwarzschild-de Sitter spacetime}

\date{\today}

\begin{abstract}
The Painlev\'e-Gullstrand coordinates are extended to describe the black hole in the cosmological environment:
the Schwarzschild-de-Sitter black hole, which has two horizons. The extension is made using the Arnowitt-Deser-Misner formalism. In this extension, which describes the metric in the whole range of radial coordinates $0<r < \infty$, there is the point $r=r_0$ at which the shift function (velocity) changes sign. At this point the observer is at rest, while the observers at $r<r_0$ are free falling to the black hole  and the observers at $r>r_0$  are free falling towards the cosmological horizon. The existence of the stationary observer allows to determine the temperature of Hawking radiation,  which is in agreement with Ref. \cite{BoussoHawking1996}. It is the red-shifted modification of the conventional Hawking temperature determined by the gravity at the horizon.
We also consider the Painlev\'e-Gullstrand coordinates and their extension for such configurations as Schwarzschild-de-Sitter white hole, where the sign of the shift function is everywhere positive;  the black hole in the environment of the contracting  de Sitter spacetime,  where the sign of the shift function is everywhere negative; and the white hole in  the contracting  de Sitter spacetime, where the shift velocity changes sign at $r=r_0$.
\end{abstract}

\maketitle


\section{Introduction}

The Painlev\'e-Gullstrand (PG) coordinates\cite{Painleve,Gullstrand} provide the specific  framework for study the black hole physics and cosmology, see recent papers \cite{Visser2022,Visser2022b,Minamitsuji2022,Volovik2022b,Hennigar2021} and references therein. These coordinates have no singularities at the horizons and thus allow for the analytic consideration both inside and outside the horizons. This is useful in particular for calculations of the Hawking temperature in the tunneling formalism.\cite{Volovik1999,Wilczek2000}

However, there are situations when the standard PG formulation does not work, see e.g. Ref. \cite{Faraoni2020}, which requires some modification of these coordinates. Typical example is the Reissner-Nordstr\"om (RN) black hole, where   the standard PG coordinates are not determined in such region inside the horizon, where the shift velocity becomes imaginary. The corresponding modification of the PG coordinates can be found in Ref.\cite{Volovik2003b} 
This modification was used  to study the thermodynamics of black holes with several horizons.\cite{Volovik2022b} The PG coordinates and their  extended  versions are also used to study the analogs of general relativity and Hawking radiation in condensed matter systems. The corresponding velocity of the free falling observer  is simulated either by the velocity of the liquid (acoustic metric),\cite{Unruh1981,Visser1998,Volovik2003} or by tilting the Weyl cone in Weyl semimetals.\cite{Volovik2016,Wilczek2020}

Here we consider the modification of the PG coordinates for the black holes in the cosmological environment. We discuss this on the particular example of the Schwarzschild--de Sitter (SdS) spacetime.\cite{BoussoHawking1996,BoussoHawking1998,Shankaranarayanan2003}

 In both cases, for the RN and SdS configurations, the extension of the PG coordinates is related to the Arnowitt-Deser-Misner (ADM)  formalism.\cite{ADM2008} This  formalism is typically used for the  Hamiltonian formulation of general relativity in terms of the Poisson brackets. The corresponding metric has the form
\begin{equation}
ds^2 = ( \gamma_{ik}N^iN^k - N^2)dt^2 +2\gamma_{ik}N^k dt dx^i  + \gamma_{ik}dx^i  dx^k\,.
\label{ADM}
\end{equation}
Here $N$ and $N^i$ are lapse and shift functions correspondingly, and $\gamma_{ik}$ are the space components of the ADM metric.
The ADM metric is similar to the acoustic metric,\cite{Unruh1981,Visser1998,Volovik2003} where the role of the shift vector $N^i$ is played by the velocity $v^i$ of the liquid: $N^i=- v^i$.
\begin{equation}
ds^2 =- N^2dt^2 + \gamma_{ik}(dx^i - v^i dt)( dx^k -v^k dt)\,.
\label{ADMacoustic}
\end{equation}

\section{Spherically symmetric stationary metrics}
\label{Holes}

Let us recall the coordinate transformations which connect the static spherically symmetric spacetime, which is typical for black and white holes in the cosmological environment, and its ADM representation.
The spherically symmetric ADM metric has the following form:
\begin{equation}
ds^2 =- N^2dt^2 +\gamma_{rr}(dr -v dt)^2+ r^2 d\Omega^2\,.
\label{ADMspherical}
\end{equation}
To connect  three functions $N$, $\gamma_{rr}$ and $v$ with the static black hole metric, the transformation $dt \rightarrow d\tilde t - q(r)dr$ is made, which gives:
\begin{eqnarray}
ds^2 =- N^2(d\tilde t - q(r)dr)^2 +
\nonumber 
\\
 \gamma_{rr}\left(dr^2  +v^2(d\tilde t - q(r)dr)^2 - 2v dr (d\tilde t - q(r)dr)\right) + r^2 d\Omega^2.
\label{ADMspherical2}
\end{eqnarray}
The cancellation of the off-diagonal terms 
\begin{equation}
0=2dt dr \left( N^2 q - \gamma_{rr}v^2q  - \gamma_{rr}v  \right) \,,
\label{off_diag}
\end{equation}
gives the following function $q(r)$:
\begin{equation}
q(r)= \frac{\gamma_{rr} v}{N^2 - v^2 \gamma_{rr}} \,,
\label{f}
\end{equation}
and the diagonal metric:
\begin{equation}
ds^2 =- (N^2- v^2\gamma_{rr})d\tilde t^2+dr^2 \gamma_{rr} \frac{N^2} {N^2- v^2\gamma_{rr}} + r^2 d\Omega^2\,.
\label{ADMdiag}
\end{equation}
If one chooses
\begin{equation}
\gamma_{rr}=\frac{1}{N^2}\,,
\label{Condition1}
\end{equation}
one obtains the fully static metric  (manifestly static spacetime in classification of Ref. \cite{Visser2022c}) describing the gravitational field of the spherical object:
\begin{equation}
ds^2= -f(r) d\tilde t^2 + \frac{dr^2}{f(r)}  + r^2d\Omega^2\,,
\label{static}
\end{equation}
where
\begin{equation}
f(r)=N^2 -\frac{v^2}{N^2} \,.
\label{g00}
\end{equation}
Eq.(\ref{g00}) contains two functions, $N$ and $v$. These two functions must match the  given static spherically symmetric solution of the Einstein equations. Then the ADM metric is obtained from this fully static metric by the coordinate transformation $dt \rightarrow d\tilde t - q(r)dr$, which is generated by the function $q(r)$ in Eq.(\ref{f}):
\begin{equation}
q(r) = \frac{v(r)}{N^4(r) - v^2(r)}\,.
\label{f_condition}
\end{equation}

The conventional Painleve-Gullstrand (PG) coordinates correspond to the particular case, when the lapse function is $N=1$:
\begin{equation}
ds^2 =- dt^2 +(dr -v(r) dt)^2+ r^2d\Omega^2\,.
\label{PGmetricl}
\end{equation}
The shift function is the velocity of the free falling observer.

We consider the extension of PG coordinates, which corresponds to the general lapse function $N(r)$.

\section{Reissner-Nordstr\"om black hole}
\label{RNHoles}

Let us consider first the extension of the PG metric on example of  the  Reissner-Nordstr\"om (RN) black hole, which has two horizons.
For the  RN black hole there is the following fully static solution of Einstein equations in static coordinates:
\begin{equation}
f(r) = 1 - \frac{2GM}{r}  +\frac{Q^2}{r^2} \,,
\label{RN}
\end{equation}
In the conventional PG coordinates (\ref{PGmetricl}), where $N=1$, the function $f(r)$ in Eq.(\ref{g00})  and the shift function $v(r)$ are given by
\begin{equation}
f(r)= 1- v^2(r) \,\,,\,\, v^2(r)=\frac{2GM}{r} - \frac{Q^2}{r^2}\,.
\label{velocityRNPG}
\end{equation} 
Such shift function $v(r)$ is not determined in the region $r<r_+ r_-/(r_+ + r_-)$, where $r_+$ and $r_-$ are correspondingly the outer and inner horizons. In this region $v^2<0$.

This coordinate problem can be treated by extension of the PG coordinates to the general case, when $N^2(r)\neq 1$. The shift function $v(r)$  becomes well defined in the whole range of distances if we choose the following lapse function $N(r)$:\cite{Volovik2003b} 
\begin{equation}
N^2(r)=1+\frac{Q^2}{r^2}\,\,, \,\,    v^2(r)=\frac{2GM}{r} \left( 1+\frac{Q^2}{r^2}\right)  \,.
\label{RNPG}
\end{equation}
Such extension allows to study the thermodynamics of the black and white holes with two horizons.\cite{Volovik2022b} 

Another example of the extension of PG coordinates to $N^2(r) \neq 1$ is represented 
by the acoustic metric.  In this case one has $N^2(r)=c(r)$, where $c(r)$ is the speed of sound, and $v(r)$ is the velocity of the liquid. This gives the effective acoustic metric in Eq.(25) of Ref. \cite{Volovik2003b} with:
\begin{equation}
f(r) =  \frac{c^2(r) -v^2(r) }{c(r)}  \,.
\label{g00acoustic}
\end{equation}
Note that the space dependent speed of light $c(r)$ in the PG coordinates was introduced by Matt Visser in the heuristic approach to the Schwarzschild geometry.\cite{Visser2004}

\section{Schwarzschild-de-Sitter black hole}
\label{SdSHoles}

\subsection{Conventional Painleve-Gullstrand metric}

The PG coordinates for the black holes in the cosmological environment  also have some problems, see e.g. Ref.\cite{Faraoni2020}. 
Let us consider first the Schwarzschild-de-Sitter (SdS) black hole. The fully static solution for the SdS black hole is:
\begin{equation}
f(r) = 1 - \frac{2GM}{r}  -H^2r^2\,.
\label{SdSmetric}
\end{equation}
Then the shift velocity  in the conventional PG coordinates (see also Refs. \cite{Visser2022b,Minamitsuji2022}) is:
\begin{equation}
v^2(r)=\frac{2GM}{r}  + H^2r^2\,.
\label{velocitySdS}
\end{equation}

This shift function is determined in the whole range of coordinates, but there is the problem with its physical  meaning, when it is interpreted as a velocity.\cite{Volovik2009}  One may expect that the sign of the velocity is positive in the region where de Sitter is dominating, i.e. $v(r) \approx  Hr$, and the free falling observer moves towards the cosmological horizon. On the other hand, the sign must be negative in the region, where the black hole metric is dominating, i.e. $v(r) \approx- \sqrt{2GM/r}$, and the free falling observer moves to the black hole. However, in the conventional PG coordinates the velocity in Eq.(\ref{velocitySdS}) may have only one sign, since $v(r)$ nowhere crosses zero.   That is why such PG coordinates are applicable to the black hole in the contracting de Sitter, where $H<0$ and $v\rightarrow -|H|r$ (the white cosmological horizon), and thus the shift velocity is everywhere negative, $v(r)<0$. This is also applicable to the white whole in the expanding de Sitter. In this case the shift velocity is positive everywhere, $v(r)>0$. 

But Eq.(\ref{velocitySdS}) does not allow to consider the vacuum state of the SdS black hole spacetime in the whole range of the coordinates.
 There is the conformal extension of the PG metric for SdS spacetime, see Eqs.(7.27)-(7.28) in Ref. \cite{Visser2022b}.
But it is not determined in the whole space. To avoid this  problem and have the possibility to study the vacuum state and the termodynamics, one should find the global coordinate system, where the shift velocity is properly determined everywhere, at $0<r<\infty$.

\subsection{Extended Painleve-Gullstrand metric}

Let us consider the following modification of the PG coordinates, in which the shift velocity for the SdS black hole is globally determined and properly changes sign:
\begin{equation}
N^2=1-C\,\,, \,\,    v^2(r)=(1-C)\left( \frac{2GM}{r}  +H^2r^2 -C\right)  \,.
\label{SdSADM}
\end{equation}
Here $C$ is the constant parameter, which we choose in such a way, that $v^2(r)$ has minimum at $v(r_0)=0$.
This gives the following values of the parameter $C$ and of the stationary point $r_0$, where the velocity is zero:
\begin{equation}
C=3(GMH)^{2/3}  \,\,, \,\, r_0^3=\frac{GM}{H^2}  \,.
\label{parameters}
\end{equation}
The spherical surface at  $r=r_0$ is the zero-mass (or zero-gravity) surface, where the dark energy mass compensates the black hole mass $M$, see e.g. Eq.(3) in Ref.\cite{Baryshev2001}.

Then the square of shift velocity becomes:
\begin{eqnarray}
  v^2(r)=(1-C)\left(\frac{2GM}{r}  +H^2r^2 -3(GMH)^{2/3}\right) =
\label{shiftSdSADM0}
  \\
=  \frac{C(1-C)}{3} \left(\frac{2r_0}{r} + \frac{r^2}{r_0^2} - 3\right)=
\label{shiftSdSADM1}
\\
= C(1-C) (r-r_0)^2 \frac{r+2r_0}{3rr_0^2}.
\label{shiftSdSADM2}
\end{eqnarray}
The shift velocity can be written in the form, which changes sign at $r=r_0$:
\begin{equation}
  v(r)= \sqrt{C(1-C)} \,\sqrt{\frac{r+2r_0}{3rr_0^2}}\,\,(r-r_0)\,.
\label{vSdSADM3}
\end{equation}
Near the stationary point  $r=r_0$ it is:
\begin{equation}
  v(r)\approx \sqrt{3(1-C)}\,H(r-r_0) \,\,,\, |r-r_0|\ll r_0\,.
\label{vSdSADM4}
\end{equation}

The velocity in Eq.(\ref{vSdSADM3}) describes the situation in which an observer at point $r=r_0$ is at rest; 
an observer at $r<r_0$ is free falling to the black hole;  and an observer at $r>r_0$  is free falling towards cosmological horizon. The observer at rest plays the same role  as an observer at infinity in the pure Schwarzschild case, who measures the temperature of the Hawking radiation. The existence of such stationary observer in SdS has been discussed in Ref.\cite{BoussoHawking1996}, where it was shown that the temperature of the Hawking radiation measured by this stationary observer is modified by the parameter $\gamma_t$. This parameter is related to the parameter  $C$ as $\gamma_t=(1-C)^{-1/2}$, see Sec.\ref{HawkingModified} below (in Ref. \cite{Shankaranarayanan2003} the parameter $\gamma_t$ is denoted as $\alpha$).

Typically the cosmological horizon radius $r_H\sim 1/H$ is much bigger that the radius of the black hole horizon $r_b\sim 2MG$,
and thus $C\ll 1$ and $r_b \ll r_0 \ll r_H$.

The function $q(r)$ in Eq.(\ref{f}), which enters the coordinate transformation,  is:
\begin{equation}
q(r)={\rm sgn}(r-r_0)\,(1-C)^{-1/2} \frac{\left( \frac{2GM}{r}  +H^2r^2 -C\right)^{1/2}}{1- \frac{2GM}{r}  -H^2r^2 }\,.
\label{SdSADMf}
\end{equation}
At $H=0$ and $C<0$ this function describes the transformation to the Martel-Poisson coordinates,\cite{MartelPoisson2001,Faraoni2020} with $p=1/(1-C)$. In general, the function $q(r)$ has one or two poles: at one horizon or at both horizons.  Near the poles Eq.(\ref{SdSADMf}) has the form
\begin{equation}
q(r)\approx \, \frac{{\rm sgn}(r-r_0)}{1- \frac{2GM}{r}  -H^2r^2 }=\frac{{\rm sgn}(r-r_0)}{f(r)}\,.
\label{polesSdSADMf}
\end{equation}

\subsection{Schwarzschild-de-Sitter with black and white horizons}

We can compare the SdS black hole with other configurations.
For the SdS white hole, where the sign of the shift function is everywhere positive, and the original PG coordinates are applicable, one has $q(r)=1/f(r)$. For the black hole in the environment of the contracting  de Sitter spacetime, $H<0$, where the sign of the shift function is everywhere negative, and the original PG coordinates are also applicable, one has $q(r)=-1/f(r)$. For the white hole in  the contracting  de Sitter spacetime the shift velocity changes sign at $r=r_0$:
\begin{equation}
    v(r)=- \sqrt{C(1-C)} \,\sqrt{\frac{r+2r_0}{3rr_0^2}}\,\,(r-r_0) \,. 
\label{WHcontractingl}
\end{equation}

In principle, the mixed configurations of static and PG coordinates are also possible. For example, the configuration  with static $v(r)=0$ spacetime at $r<r_0$  and the extended PG metric at $r>r_0$ is given by:
\begin{eqnarray}
r<r_0\, :  \,\, N^2=f(r)  \,\,,\,  \,\,   v(r) = 0 \, ;
\label{mixed2}
\\
 r>r_0\, :  \,\,   N^2=1-C\,,
  \nonumber
\\
  v(r)= \sqrt{C(1-C)} \,\sqrt{\frac{r+2r_0}{3rr_0^2}}\,\,(r-r_0) \,.
 \label{mixed1}
 \end{eqnarray}
 The shift function $v(r)$ and lapse function $N^2$ are continuous across $r=r_0$.
 
If so, there are 9 possible coordinate frames which can be obtained from each other by singular coordinate transformations. Since in each frame the coordinates are well determined in the whole range $0<r<\infty$, one may suggest that in each frame one can determine the physical state which is characterized by its entropy and temperature. In principle, one may have 9 realizations of the SdS structure:

1. PG black hole in expanding de Sitter

2. PG white hole in expanding de Sitter

3. PG black hole in contracting de Sitter

4. PG white hole in contracting de Sitter

5. The fully static configuration

6. Static hole in expanding de Sitter

7. Static hole in contracting de Sitter

8. PG black hole in static de Sitter

9. PG white hole in static de Sitter

The important point is that the transformations between these 9 frames have singularities at the horizons, where the function $q(r)$ has poles. While the configurations, which are connected by the regular coordinate transformations, are equivalent according to the diffeomorphism covariance, one may expect  that the configurations connected by singular transformations are physically nonequivalent. They have different vacuum states and different thermodynamics. 
This issue is similar to the problem of conformally connected frames, the Jordan and the Einstein
frames in the scalar-tensor gravity, which in principle may correspond to physically different situations, see Ref. \cite{Faraoni1999}  and recent review \cite{Bhattacharya2022}.

Example is provided by the PG coordinates for black and white holes in the flat space. These states are physically different, but can be obtained from each other by the singular coordinate transformation.\cite{Volovik2022b} 
This coordinate transformation is used to obtain the semiclassical tunneling rate of the quantum transition from the black hole state  to the state of the white hole. The tunneling exponent allows to calculate the entropy of the white hole, which differs from the entropy of black hole by sign, $S_{\rm WH}=-S_{\rm BH}$.\cite{Volovik2022b} Such quantum transition between the black and white holes may take place at the final stage of the black hole history, see discussions in Refs.\cite{Barcelo2014,Barcelo2017,Rovelli2019,Rovelli2018b,Uzan2020b,Uzan2020c,Uzan2020d,Bodendorfer2019,Rovelli2021a}.

Moreover, the tunneling approach using singular coordinate transformation also allows us to introduce the neutral object, which is intermediate between the black hole and the white hole. Such the fully static object  has zero entropy and zero temperature, and it does not experience the Hawking radiation. This object is described by the static Schwarzschild metric, which is be obtained by singular coordinate transformation either from the PG  black hole coordinates or from the PG white hole frame. The physical state of this fully static object is obtained by the quantum tunneling form the PG black hole state, while the further quantum tunneling transforms this object to the white hole state.

Whether the coordinate transformations between 9 realizations of SdS configurations may provide the information on the thermodynamics of these composite objects is an open question, as well as the other general problems related to the dS and SdS thermodynamics, see e.g. recent discussion in Refs.\cite{Jacobson2022b,Jacobson2022}. 
In particular, it was suggested that the Bunch-Davis vacuum\cite{BunchDavies1978} with Hawking temperature $T_c=H/2\pi$  of the cosmological horizon is the property of the de Sitter spacetime in the PG coordinate system.\cite{Parikh2006} Assuming that this is so, one may suggest that the de Sitter spacetime corresponding to the contracting  PG coordinate system, $H<0$, has negative temperature of the cosmological horizon, $T_c=-|H|/2\pi$ (see recent review on thermodynamics of systems with negative temperature \cite{Baldovin2021}).
However, this requires the proper justification, as the other conjectures related to the de Sitter spacetime and the cosmological constant problem. In particular, in the $q$-theory describing the vacuum energy, the Big Bang can be considered as the topological quantum transition between the contracting and expanding de Sitter spacetime.\cite{KlinkhamerVolovik2022} This transition is governed by the Planck scale cosmological constant, which later decays to the present value.

\subsection{Modified Hawking radiation}
\label{HawkingModified}

The energy spectrum of massless particles in the extended PG metric of SdS spacetime is given by the contravariant components of the metric:
\begin{equation}
g^{\mu\nu}p_\mu p_\nu =-\frac{1}{N^2}(E+p_r v)^2 +N^2 p_r^2 +p_\perp^2=0\,.
\label{SdSADMg}
\end{equation}
The radial trajectory of massless particle with energy $E$ is:
\begin{equation}
p_r(r,E)= -\frac{E}{N^2-v(r) }\,,
\label{pSdS}
\end{equation}
which near the poles is expressed in terms of the static metric:
\begin{equation}
p_r(r,E) \approx - \frac{2E}{f(r)}\,,
\label{pSdS2}
\end{equation}
At first glance this leads to the conventional temperature of Hawking radiation, which is determined by gravity at the horizon. However, we must take into account, that now the stationary observer is not at infinity, but is at $r=r_0$. That is why the frequency measured by the stationary observer  is red-shifted by the factor $N=(1-C)^{1/2}=1/\gamma_t$, and the Hawking temperatures of black hole and cosmological horizons are modified in agreement with Ref. \cite{BoussoHawking1996}:
\begin{equation}
T_b=\frac{T_{b0}}{N}\,\,,\,\, T_c=\frac{T_{c0}}{N} \,.
\label{SdSHawking}
\end{equation}
Here $T_{b0}$ and $T_{c0}$ are the original  Hawking temperatures of black hole and cosmological horizons determined by gravity at the horizon. 

\subsection{Colliding horizons and Schwarzschild-de-Sitter spacetime without horizons}

The global temperature of the SdS system can be determined when the black hole horizon approaches the cosmological horizon, $r_b \rightarrow r_0-0$ and  $r_c \rightarrow r_0 + 0$. But this takes place in the limit when $C\rightarrow 1$ and both temperatures approach infinity.
This adds to the general problems related to the dS and SdS thermodynamics.\cite{Jacobson2022b,Jacobson2022}.

For $C=1$ the black hole and cosmological horizons collide. At $C>1$ one has $f(r)<0$ in the whole space, and thus there are no horizons. In this case the extended PG coordinates have the lapse function $N^2=1-C <0$ and the shift velocity, which changes sign at $r=r_0$:
\begin{equation}
  v(r)= \sqrt{C(C-1)} \,\sqrt{\frac{r+2r_0}{3rr_0^2}}\,\,(r-r_0)  \,. 
\label{horizonless}
\end{equation}

\subsection{Schwarzschild-anti-de-Sitter black hole}
\label{SAdSHoles}

For the Schwarzschild-anti-de-Sitter (SAdS) black hole one has:
\begin{equation}
f(r)= 1 - \frac{2GM}{r}  + H^2r^2\,,
\label{SadSmetric}
\end{equation}

The ADM coordinates, which modify the PG coordinates for SAdS black hole, can be obtained in the same way as
for RN black hole,\cite{Volovik2003b} i.e. by substitution $Q^2/r^2 \rightarrow H^2 r^2$ in Eq.(\ref{RNPG}):
\begin{equation}
N^2=1+H^2r^2\,\,, \,\,    v^2=\frac{2GM}{r} \left( 1+H^2r^2\right)  \,.
\label{SAdSADM}
\end{equation}
The function $q(r)$ in Eq.(\ref{f_condition}) is:
\begin{equation}
q (r)= \frac{v}{N^4 - v^2}=\sqrt{\frac{2MG}{r(1+H^2r^2)}}\,
\frac{1}{1-\frac{2MG}{r} +H^2r^2}\,.
\label{f_SAdS}
\end{equation}

Hawking radiation is obtained using modification of trajectory in Eq.(46) of Ref.\cite{Volovik2022b} to apply it to the SAdS case:
\begin{equation}
p_r(r,E)= \frac{E}{v(r) -(1+H^2r^2)}\,,
\label{p_SAdS}
\end{equation}
which near the pole has the form
\begin{equation}
p_r(r,E) \approx - \frac{2E}{1-\frac{2MG}{r} +H^2r^2}=-\frac{2E}{f(r)} \,.
\label{p_SAdS2}
\end{equation}
This trajectory gives the conventional Hawking temperature, which is determined by gravity at the horizon. But there is no stationary observer, who could measure such temperature, and thus there is no global temperature. 

\section{Conclusion}

In conclusion, we extended the Painlev\'e-Gullstrand coordinates to the Schwarzschild-de-Sitter black hole, using the Arnowitt-Deser-Misner  formalism. The extended Painlev\'e-Gullstrand coordinates describe the metric in the whole range of radial coordinates, $0<r < \infty$. The extended Painlev\'e-Gullstrand metric contains the point $r=r_0$, at which the shift function changes sign. At this point the observer is at rest, which allows to determine the temperature of Hawking radiation. The observers at $r<r_0$ are free falling to the black hole, while  the observers at $r>r_0$  are free falling towards the cosmological horizon. The temperature of Hawking radiation,  which is measured by the stationary observer,  is the red-shifted modification of the conventional Hawking temperature determined by the gravity at the horizon. This is in agreement with results of Ref. \cite{BoussoHawking1996}. 

The extension of the  Painlev\'e-Gullstrand coordinates suggests the existence of nine physically different Schwarzschild-de-Sitter objects. The next step would be  to obtain the thermodynamic properties of these objects using the quantum tunneling between nine cosmological objects, which rate is determined by the singular coordinate transformations. 

 {\bf Acknowledgements}.  This work has been supported by the European Research Council (ERC) under the European Union's Horizon 2020 research and innovation programme (Grant Agreement No. 694248).

\end{document}